# Interactive tutorial to improve student understanding of single photon experiments involving a Mach-Zehnder Interferometer

Emily Marshman and Chandralekha Singh

*Department of Physics and Astronomy, University of Pittsburgh, Pittsburgh, PA, 15260, USA*

**Abstract:** We have developed and evaluated a Quantum Interactive Learning Tutorial (QuILT) on a Mach-Zehnder Interferometer with single photons to expose upper-level students in quantum mechanics courses to contemporary quantum optics applications. The QuILT strives to help students develop the ability to apply fundamental quantum principles to physical situations in quantum optics and explore the differences between classical and quantum ideas. The QuILT adapts visualization tools to help students build physical intuition about counter-intuitive quantum optics phenomena with single photons including a quantum eraser setup and focuses on helping them integrate qualitative and quantitative understanding. We discuss findings from in-class evaluations.

## I. INTRODUCTION

Prior studies have examined students' understanding of classical optics, e.g., in the context of their understanding of geometrical optics and vision [1] and improving student understanding. However, few investigations have focused on helping students learn quantum optics better. Quantum mechanics, in general, can be a challenging subject for students partly because it is unintuitive and abstract [2-16]. A quantum optics experiment which has been conducted in undergraduate laboratories to illustrate fundamental principles of quantum mechanics involves the Mach-Zehnder Interferometer (MZI) with single photons [17-23]. We have developed and evaluated a quantum interactive learning tutorial (QuILT) using gedanken (thought) experiments and simulations involving a MZI to help students learn about single photon interference. The QuILT focuses on using this quantum optics experiment to help students learn topics such as the wave-particle duality of a single photon, interference of a single photon with itself, and probabilistic nature of quantum measurements. Students also learn how adding photo-detectors and optical elements such as beam-splitters and polarizers in the paths of the MZI affect the measurement outcomes.

Figure 1 shows the MZI setup. For simplicity, the following assumptions are made: 1) all optical elements are ideal; 2) the non-polarizing beam-splitters (BS1 and BS2) are infinitesimally thin such that there is no phase shift when a single photon propagates through them; 3) the monochromatic single photons travel the same distance in vacuum in the upper path (U) and lower path (L) of the MZI; and 4) the initial MZI without the phase shifter is set up such that there is completely constructive interference at photo-detector 1 (D1) and completely destructive interference at photo-detector 2 (D2).

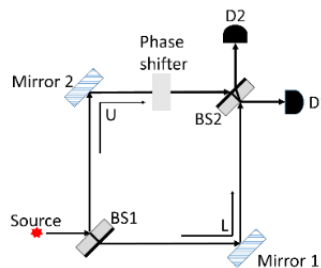

*Figure 1. MZI setup with a phase shifter in the U path*

If single photons are emitted from the source in Figure 1, BS1 causes each single photon to be in a superposition state of the path states U and L. The photon path states reflect off of the mirrors and recombine in beam-splitter BS2. BS2 mixes the photon path states such that each component of the photon state along the U and L paths can be projected into the photo-detectors D1 and D2 in Figure 1. The projection of both components leads to interference at the photo-detectors (called detectors from now on). Depending on the thickness of the phase shifter, interference observed at detectors D1 and D2 can be constructive, destructive, or intermediate. Observing interference of a single photon with itself at D1 and D2 can be interpreted in terms of not having "which-path" information (WPI) about the single photon [17-23]. WPI is a common terminology associated with these types of experiments popularized by

Wheeler [24]. WPI is unknown (as in the setup shown in Fig. 1) if both components of the photon state can be projected into D1 and D2 and the projection of both components at each detector leads to interference. When WPI is unknown and a large number of single photons are sent through the setup, if a phase shifter is inserted in one of the paths of the MZI (as in the U path in Fig. 1) and its thickness is varied, the probability of the photons arriving at D1 and D2 will change with the thickness of the phase shifter due to interference of the components of the single photon state from the U and L paths.

In a simplified quantum mechanical model of a photon state which accounts for the two paths $U$ and $L$ (see Fig. 1), a single photon traveling through the MZI can be considered to be a two state quantum system. If a basis is chosen in which the state of the photon in the upper state is denoted by $|U\rangle = \begin{pmatrix}1\\0\end{pmatrix}$ and the state of the photon in the lower state is denoted by $|L\rangle = \begin{pmatrix}0\\1\end{pmatrix}$ (and we arbitrarily denote the initial state of the photon emitted from the source as $|I\rangle = \begin{pmatrix}1\\0\end{pmatrix}$, the state of the photon propagating towards detector D1 as path state $|D1\rangle = \begin{pmatrix}1\\0\end{pmatrix}$, and the state of the photon propagating towards detector D2 as the path state $|D2\rangle = \begin{pmatrix}0\\1\end{pmatrix}$), the matrix representations of the quantum mechanical operators that correspond to beam-splitter 1 $[BS1]$, beam-splitter 2 $[BS2]$, the mirrors $[M]$, and a phase shifter in the upper path $[PS_U]$ when the basis vectors are chosen in the order $|U\rangle, |L\rangle$ are: $[BS1] = \frac{1}{\sqrt{2}}\begin{bmatrix}-1 & 1\\1 & 1\end{bmatrix}$, $[BS2] = \frac{1}{\sqrt{2}}\begin{bmatrix}1 & -1\\1 & 1\end{bmatrix}$, $[M] = \begin{bmatrix}-1 & 0\\0 & -1\end{bmatrix}$, and $[PS_U] = \begin{bmatrix}e^{i\phi_{PS}} & 0\\0 & 1\end{bmatrix}$, where $\phi_{PS}$ is the phase shift introduced by the phase shifter.

The final state of a photon $|F\rangle$ in Figure 1 can be determined by operating on the initial photon state with the operators corresponding to the optical elements in the appropriate time-ordered manner: $|F\rangle = [BS2][PS_U][M][BS1]|I\rangle = \frac{1}{2}\begin{pmatrix}e^{i\phi_{PS}}+1\\e^{i\phi_{PS}}-1\end{pmatrix}$. The probability of detector D1 registering a photon is $|\langle D1|F\rangle|^2 = (1+\cos\phi_{PS})/2$ (the probability of detector D2 registering a photon is $|\langle D2|F\rangle|^2 = (1-\cos\phi_{PS})/2$). Since the probability of a detector registering a photon depends on the phase shift of the phase shifter, interference effects are observed when the phase shift of the phase shifter is gradually changed. If there is no phase shifter in the upper path in Fig. 1 ($\phi_{PS} = 0$), all photons are registered at detector D1 ($|F\rangle = |D1\rangle = \begin{pmatrix}1\\0\end{pmatrix}$) since completely constructive interference takes place at detector D1 and completely destructive interference takes place at detector D2.

On the other hand, if the components of the photon path state are not recombined, there is no possibility for interference of the photon path states to occur at the detectors. In this case, WPI is known about a photon that arrives at a detector D1 or D2. In other words, WPI is "known" about a photon if only one component of the photon path state can be projected into each detector. For example, if BS2 is removed from the setup (see Fig. 2), WPI is known for all single photons arriving at the detectors because only the component of a photon state along the U path can be projected in D1 and only the component of a photon state along the L path can be projected in D2. When WPI is known, each detector (D1 and D2) has an equal probability of clicking. A detector clicks when a photon is detected by it and is absorbed (the state of the single photon collapses, i.e., the single photon state is no longer in a superposition of the U and L path states). However, when WPI is known, there is no way to know a priori which detector will click when a photon is emitted until the photon state collapses either at D1 or at D2 with equal likelihood. When WPI is known, changing the thickness of a phase shifter in one of the paths will not affect the probability of each detector clicking when photons are registered (equal probability for all thicknesses of the phase shifter).

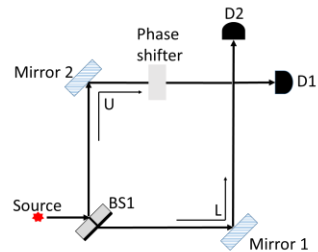

*Figure 2. MZI setup with beam-splitter 2 (BS2) removed*

If beam-splitter 2 is removed (see Fig. 2), the final state of the photon is $|F\rangle = [PS_U][M][BS1]|I\rangle = \frac{1}{\sqrt{2}}\begin{pmatrix}e^{i\phi_{PS}}\\-1\end{pmatrix}$. The probability of detector D1 registering a photon is $|\langle D1|F\rangle|^2 = 1/2$ (the probability of detector D2 registering a photon

is also 1/2). Thus, the probability of the detectors registering a photon does not depend on the phase shift of the phase shifter and interference effects are not observed when the phase shift of the phase shifter is gradually changed.

When polarizers are added to the MZI setup, they can affect (and even eliminate or reinstate) the interference of a single photon with itself at the detectors [18,21-23]. In all the MZI setups discussed, it is assumed that the detectors are polarization sensitive and the single photons are linearly polarized. In Figure 3, two orthogonal polarizers are placed in the U and L paths of the MZI. If the source emits a large number ($N$) of +45° polarized single photons, $N/2$ photons are absorbed by the polarizers. If a detector in Fig. 3 measures a vertically polarized photon, only one component of the photon path state can be projected in the detector (i.e., the L path state) and WPI is known. If a detector measures a horizontally polarized photon, again, only one component of the photon path state can be projected in the detector (i.e., the U path state) and WPI is known. WPI is known for all photons arriving at the detectors, and there is an equal probability of each detector registering a photon ($N/4$ photons arrive at each detector). There is no interference observed at the detectors. Inserting a phase shifter and changing its thickness gradually will not affect the number of photons arriving at the detectors in Fig. 3.

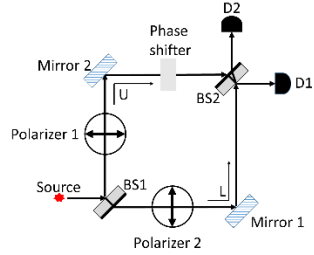

*Figure 3. MZI setup with a polarizer with a vertical transmission axis placed in the U path and a polarizer with a horizontal transmission axis placed in the L path.*

If we only consider photon polarization states, the polarization state of a vertically polarized photon can be denoted by $|V\rangle = \begin{pmatrix}1\\0\end{pmatrix}$ and the polarization state of a horizontally polarized photon can be denoted by $|H\rangle = \begin{pmatrix}0\\1\end{pmatrix}$. These two polarizations are linearly independent and all other photon polarizations can be constructed from these states, e.g., $|+45°\rangle = (|V\rangle + |H\rangle)/\sqrt{2}$. The Hilbert space involving both path states and polarization states is a product space. The product space of the polarization states $|V\rangle$ and $|H\rangle$ and the path states $|U\rangle$ and $|L\rangle$ is four dimensional, and the basis vectors are $|U\rangle \otimes |V\rangle = |UV\rangle$, $|U\rangle \otimes |H\rangle = |UH\rangle$, $|L\rangle \otimes |V\rangle = |LV\rangle$, $|L\rangle \otimes |H\rangle = |LH\rangle$. If the initial path state of the photon emitted from the source is denoted by $|I\rangle = \begin{pmatrix}1\\0\end{pmatrix}$ and the initial polarization state of the photon is $|+45°\rangle = (|V\rangle + |H\rangle)/\sqrt{2}$, in the $4 \times 4$ product space, the initial state of the photon $|I_{45°}\rangle$ is $|I_{45°}\rangle = |I\rangle \otimes |45°\rangle = \frac{1}{\sqrt{2}}\begin{pmatrix}1\\1\\0\\0\end{pmatrix}$. The matrix representations of the quantum mechanical operators that correspond to beam-splitter 1 $[BS1]$, beam-splitter 2 $[BS2]$, the mirrors $[M]$, a phase shifter in the upper path $[PS_U]$, a horizontal polarizer in the upper path $[P_{UH}]$, a vertical polarizer in the lower path $[P_{LV}]$, and a +45 polarizer in the path between BS2 and detector D1 $[P_{D1,+45°}]$ when the basis vectors are chosen in the order $|UV\rangle$, $|UH\rangle$, $|LV\rangle$, $|LH\rangle$ are: $[BS1] = \frac{1}{\sqrt{2}}\begin{bmatrix}-1 & 0 & 1 & 0\\0 & -1 & 0 & 1\\1 & 0 & 1 & 0\\0 & 1 & 0 & 1\end{bmatrix}$, $[BS2] = \frac{1}{\sqrt{2}}\begin{bmatrix}1 & 0 & -1 & 0\\0 & 1 & 0 & -1\\1 & 0 & 1 & 0\\0 & 1 & 0 & 1\end{bmatrix}$, $[M] = -\begin{bmatrix}1 & 0 & 0 & 0\\0 & 1 & 0 & 0\\0 & 0 & 1 & 0\\0 & 0 & 0 & 1\end{bmatrix} = -\hat{I}$, $[PS_U] = \begin{bmatrix}e^{i\varphi_{ps}} & 0 & 0 & 0\\0 & e^{i\varphi_{ps}} & 0 & 0\\0 & 0 & 1 & 0\\0 & 0 & 0 & 1\end{bmatrix}$, where $\phi_{PS}$ is the phase shift introduced by the phase shifter, $[P_{UH}] = \begin{bmatrix}0 & 0 & 0 & 0\\0 & 1 & 0 & 0\\0 & 0 & 1 & 0\\0 & 0 & 0 & 1\end{bmatrix}$, and $[P_{LV}] = \begin{bmatrix}1 & 0 & 0 & 0\\0 & 1 & 0 & 0\\0 & 0 & 1 & 0\\0 & 0 & 0 & 0\end{bmatrix}$.

The final state of a photon $|F\rangle$ in Figure 3 can be determined by operating on the initial photon state with the operators corresponding to the optical elements in the appropriate time-ordered manner: $|F\rangle = [BS2][PS_U][M][P_{LV}][P_{UH}][BS1]|I_{45°}\rangle = (|UV\rangle + e^{i\varphi_{PS}}|UH\rangle - |LV\rangle + e^{i\varphi_{PS}}|LH\rangle)/(2\sqrt{2})$. The probability of detector D1 registering a horizontally polarized photon is $|e^{i\varphi_{PS}}/(2\sqrt{2})|^2 = 1/8$ and the probability of detector D1 registering a vertically polarized photon is $|1/(2\sqrt{2})|^2 = 1/8$. The total probability of detector D1 registering a photon is $1/8 + 1/8 = 1/4$. The total probability of detector D2 registering a photon is also $1/4$. Thus, in the case shown in Figure 3, the probability of a detector registering a horizontally or vertically polarized photon does not depend on the phase shift of the phase shifter and interference effects are not observed when the phase shift of the phase shifter is gradually changed.

Figure 4 shows a quantum eraser setup in which two orthogonal polarizers are placed in the two paths of the MZI and a third polarizer is placed between BS2 and detector D1. The third polarizer has a transmission axis which is different from the two orthogonal polarizers. Without polarizer 3, WPI is known for all photons arriving at the detectors (as in Figure 3) and interference is not observed at the detectors. However, when polarizer 3 is inserted between BS2 and detector D1, both the U and L path states are projected into D1 and WPI is unknown for all photons. For example, if detector D1 measures vertically polarized photons, both components of the photon path state are projected into detector D1 and WPI is unknown. Similarly, if D1 measures horizontally polarized photons, both components of the photon path state are projected into detector D1 and WPI is again unknown. Interference is observed at detector D1. If a phase shifter is inserted into one of the paths of the MZI, changing its thickness gradually will change the number of photons arriving at D1. Because polarizer 3 eliminates WPI at the detector D1, this MZI setup is called a quantum eraser. However, in Fig. 4, WPI is known at detector D2 and no interference is observed there. Inserting a phase shifter into one of the paths of the MZI and changing its thickness gradually will not affect the number of photons that arrive at D2.

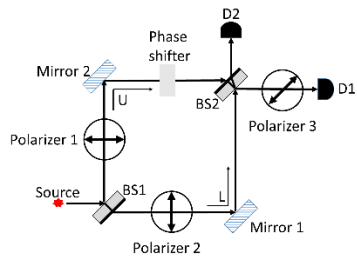

*Figure 4. Quantum eraser setup*

The matrix representing a third polarizer with a $+45°$ polarization axis inserted between beam-splitter 2 and detector D1, $[P_{D1,+45°}]$ (as shown in Figure 4) is $[P_{D1,+45°}] = \begin{bmatrix} \frac{1}{2} & \frac{1}{2} & 0 & 0 \\ \frac{1}{2} & \frac{1}{2} & 0 & 0 \\ 0 & 0 & 1 & 0 \\ 0 & 0 & 0 & 1 \end{bmatrix}$. The final state of the photon in Figure 4 can be determined by operating on the initial photon state with the operators corresponding to optical elements in the appropriate time-ordered manner: $|F\rangle = [P_{D1,+45°}][BS2][PS_U][M][P_{LV}][P_{UH}][BS1]|I_{45°}\rangle =$
$((\frac{1}{2} + \frac{1}{2}e^{i\varphi_{PS}})|UV\rangle + (\frac{1}{2} + \frac{1}{2}e^{i\varphi_{PS}})|UH\rangle - |LV\rangle + e^{i\varphi_{PS}}|LH\rangle)/(2\sqrt{2})$. The probability of detector D1 registering a horizontally polarized photon is $\left|(\frac{1}{2} + \frac{1}{2}e^{i\varphi_{PS}})/(2\sqrt{2})\right|^2 = (1 + \cos\varphi_{PS})/16$ and the probability of detector D1 registering a vertically polarized photon is $\left|(\frac{1}{2} + \frac{1}{2}e^{i\varphi_{PS}})/(2\sqrt{2})\right|^2 = (1 + \cos\varphi_{PS})/16$ (the total probability of detector D1 registering a photon is $(1 + \cos\varphi_{PS})/8$). In the quantum eraser case shown in Figure 4, the probability of detector D1 registering a horizontally or vertically polarized photon depends on the phase shift of the phase shifter and interference effects are observed when the phase shift of the phase shifter is gradually changed. WPI is unknown at detector D1.

The quantum eraser setup also distinguishes between a stream of unpolarized photons and photons which have been polarized at $+45°$. If the source emits unpolarized photons, one can consider half of the photons emitted to be vertically polarized and half of the photons emitted to be horizontally polarized (or half of the photons emitted can be considered $+45°$ polarized and half of the photons $-45°$ polarized). In Fig. 4, if one considers unpolarized photons as a mixture of half vertically polarized and half horizontally polarized photons incident at BS1 randomly, a single photon with horizontal polarization can only go through the upper path and a single photon with a vertical polarization can only go through the lower path. If the photon passes through polarizer 3, each detector can only project one component of the photon path state and WPI is known. Interference effects are not observed. Inserting a phase shifter and changing its thickness gradually will not affect the number of photons arriving at the detectors. On the other hand, In Fig. 4, if one considers unpolarized photons as a mixture of half of the photons polarized at $+45°$ and half of the photons polarized at $-45°$ incident at BS1 randomly, the total probability of unpolarized photons arriving at detector

D1 can be determined by averaging the total probabilities of detector D1 registering a photon for the two cases in which the source emits +45° single photons and -45° single photons. In the case in which the source emits +45° single photons, the total probability of detector D1 registering a photon is $(1 + \cos \varphi_{PS})/8$. In the case in which the source emits -45° single photons, the total probability of detector D1 registering a photon is $(1 - \cos \varphi_{PS})/8$. The average of these two probabilities ($(1 + \cos \varphi_{PS})/8$ and $(1 - \cos \varphi_{PS})/8$) is 1/8, indicating that for unpolarized light (which can be treated as a mixture in which half of the photons are +45° polarized and half of the photons are -45° polarized) the setup in Fig. 4 does not erase which path information and changing the phase shift of the phase shifter does not affect the number of photons arriving at the detector D1. However, in the quantum eraser setup (see Fig. 4), if the source emits a stream of +45° polarized single photons, both components of the photon path state can be projected in detector D1. The total probability of detector D1 registering a photon is $(1 + \cos \varphi_{PS})/8$ and depends on the phase shift of the phase shifter. Interference effects are observed at detector D1. Thus, the quantum eraser distinguishes between a stream of unpolarized photons and photons which have been polarized at +45°. In the quantum eraser setup, interference effects will not be observed at detector D1 when unpolarized photons are emitted and interference effects are observed at detector D1 when polarized photons are emitted.

## II. METHODOLOGY FOR THE INVESTIGATION OF STUDENT DIFFICULTIES

Student difficulties involving the MZI with single photons were investigated by administering open-ended questions to upper-level undergraduate and Ph.D. students in physics and conducting individual interviews with 15 students in quantum mechanics courses after traditional instruction in relevant concepts. The traditional instruction included an overview of the MZI setup and students learned about phase differences, reflection off of mirrors, propagation of light through the beam-splitters, and the meaning of what happens when the detectors "click." The open-ended questions were graded using rubrics which were developed by the two investigators together. A subset of the questions was graded separately by the investigators. After comparing the grading, the investigators discussed any disagreements in grading and resolved them. The final inter-rater reliability in the grading of open-ended questions is better than 90%.

We conducted 15 individual interviews, which used a semi-structured, think-aloud protocol [25], to better understand the rationale for student responses before, during, and after the development of different versions of the MZI tutorial and the corresponding pretest and posttest. During the semi-structured interviews, upper-level undergraduates and Ph.D. students were asked to verbalize their thoughts while they answered questions. Students read the questions related to the MZI setup and answered them to the best of their ability without being disturbed. We prompted them to think aloud if they were quiet for a long time. After students had finished answering a particular question to the best of their ability, we asked them to further clarify and elaborate issues that they had not clearly addressed earlier.

## III. STUDENT DIFFICULTIES

During the preliminary development of the QuILT, we investigated the difficulties students have with the relevant concepts including the wave-particle duality of a photon, interference of a single photon with itself, and the probabilistic nature of quantum measurements in order to effectively address them. Some of the common difficulties found in the interviews included students struggling with the interference of a classical beam of light through the MZI, ignoring the wave nature of single photons, claiming that a photon is split into two photons after BS1 (see Fig. 1), difficulty with how BS2 affects measurement outcomes, difficulty with how polarizers can act as partial measurement devices and alter the state of a photon, and how WPI can be erased, e.g., by introducing polarizer 3 in Fig. 4.

**Difficulty with the interference of light waves at detectors after passing through the MZI:** Interviews suggest that many students did not take into account the interference phenomenon of a classical beam of light. For example, regarding a beam of light with intensity $I$ propagating through the setup shown in Fig. 1 without the phase shifter, one student stated: "There will be billions of photon[s] in one beam so… approximately half go through U and half go through L. When going through BS2 they also have equal chance to reach D1 [and] D2. So the [intensity] on each [detector] will be $I/2$." Similar to this student, other students in interviews invoked the concept of photons when reasoning about a classical beam of light. Further probing indicates that students with these types of responses had some idea that a beam of light can be treated as a stream of photons but they often failed to invoke the wave nature of light which would lead, e.g., to constructive interference at D1 and destructive interference at D2 for the setup given without a phase shifter.

In addition to the difficulties involving the intensity of a light beam through the MZI, students also had difficulty reasoning about a large number of single photons emitted from the source and how the single photons would propagate through the MZI. Students were asked to explain whether they agreed or disagreed with the following statement for the setup shown in Fig. 1 without the phase shifter and why: "If the source emits $N$ photons one at a time, the number of photons reaching detectors D1 and D2 will be $N/2$ each." Many students incorrectly agreed with this statement. For example, one student stated, "I agree because the photon has equal probability of reflecting or transmitting when it hits the beam-splitter." Students with this type of response had difficulty reasoning about how the beam-splitter causes a photon to be in a superposition of the U and L path states. They also did not take into account the phase shifts of each photon path component and how the phase difference between the U and L paths causes constructive and destructive interference of single photons at the detectors.

**Difficulties due to a single photon as a point particle model:** Students struggled with the concept of the wave/particle duality of a single photon and the fact that interference can be observed at the detectors due to a single photon state with contributions from the two paths (e.g., in Fig. 1, the photon state is in a superposition of the U and L path states after BS1 which can interfere at the detectors D1 and D2). Students often treated a single photon as a point particle, ignoring its wave-like nature. Some students claimed that a single photon can be split into two photons and it is these two photons that interfere at the detectors (instead of the fact that interference is due to the wave nature of single photons). For example, one student said "it seems like [each photon with half of the energy of the incoming photon traveling along the U and L paths of the MZI is] the only way for a photon to interfere with itself and have some probability of going through either path until getting measured." Other students claimed that neither the photon nor its energy will be split in half after BS1, but that each photon is localized in either the U or L path. These types of reasoning difficulties indicate that students struggled with the fact that a single photon can behave as a wave passing through the MZI and be in a superposition of U and L path states until a measurement is performed, e.g., at the detectors D1 and D2, and the state collapses.

**Difficulty with the role of the beam-splitter BS2:** Several students incorrectly claimed that either removing or inserting BS2 will not change the probability of the single photons arriving at each detector. For example, one student supplemented his claim as follows: "I don't see how BS2 affects/causes any asymmetry to make probabilities D1≠D2 or how BS2 causes a loss of photons." Another student who made similar claims about what happens at the detectors with and without BS2 stated, "I say still 50% each since it's symmetric." Students who treated a single photon as a point particle and ignored its wave nature did not take into account the phase shifts affecting the components of the photon state along the U and L paths due to BS1 and BS2 (e.g., in Fig. 1) which influence the interference of the single photons at the detectors D1 and D2.

**Difficulty with how a detector placed in the U or L path affects the single photon state:** Students often asserted that inserting an additional detector in the U or L path of the MZI would not affect the interference at the detectors D1 and D2 at the end (see Fig. 5). They had difficulty with the fact that an additional detector, e.g., in the L path of the MZI in Figure 5, would collapse the state of the photon to the U or L path state so that the detectors D1 or D2 after BS2 would click with equal probability and the interference would be destroyed. Instead, many students claimed that the photon state would remain delocalized in a superposition of the U and L path states (as in Fig. 1) and interference would be observed at D1 and D2 even in the situation in Fig. 5. Some students correctly stated that a detector placed in the L path would absorb some photons but incorrectly inferred that there would still be interference displayed by the photons that are not absorbed earlier and reach detector D1 and/or D2. For example, one student said "Now path L is blocked [by a detector in the L path], so only ½ as many photons should hit the [detector D1 or D2 at the end]. I don't see how there can be any but constructive interference since path lengths are the same." Further probing of students with these types of responses suggests that they struggled with how placing a detector in the U or L path amounts to a measurement and destroys the delocalized single photon state which was in a superposition of the U and L path states before the measurement.

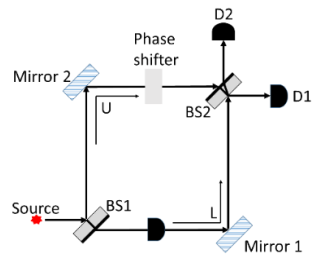

*Figure 5. MZI setup with an additional detector placed in the L path*

**Difficulty with two orthogonal polarizers placed in the U and L paths of the MZI:** Students often incorrectly claimed that interference would be displayed when two orthogonal polarizers are placed in the two paths of the MZI (see Fig. 3). As noted earlier, WPI is known about all the photons in Fig. 3, interference is destroyed, and the detectors register photons with equal probability. Many students stated that "less photons would reach the [detectors]" but that interference would still be displayed. For example, one student stated that "50% [of the photons emitted by the source display interference because] we don't measure anything until the photons hit the [detector D1] so their state vector doesn't collapse until then." These students struggled with the fact that two orthogonal polarizers placed in the two paths of the MZI correspond to a measurement of photon polarization such that either the photon gets absorbed by the polarizer or the photon with a vertical polarization that reaches D1 or D2 came only from the MZI path with the vertical polarizer and the photon with a horizontal polarization that reaches D1 or D2 came only from the path with the horizontal polarizer. Thus, WPI is known about all photons that reach D1 and D2 and no interference is observed. Students had difficulty with the fact that once a photon reaches the polarizers, the measurement of polarization collapses the state of the photon such that if a detector registers a photon with a horizontal polarization, it must have come from the U path and if a detector registers a photon with a vertical polarization, it must have come from the L path.

**Difficulty with how a quantum eraser setup in Fig. 4 erases WPI and restores interference of single photons at detector D1**: In contrast to the MZI setup with two orthogonal polarizers in which WPI is known about all photons regardless of whether they are initially polarized or unpolarized, the addition of a third polarizer as shown in Fig. 4 causes both components of the photon path state to be projected into D1, erasing WPI about the +45° polarized single photons arriving at D1 as discussed earlier (but not of unpolarized single photons). If a phase shifter is inserted in one of the paths of the MZI and its thickness is gradually changed, the interference displayed at D1 will change. Some students incorrectly claimed that the quantum eraser setup (see Fig. 4) is not different from the setup in which two orthogonal polarizers are placed in the U and L paths (see Fig. 3) except fewer photons would reach D1 because some will be absorbed by polarizer 3. Moreover, many students could not articulate why the quantum eraser setup shows interference effects at detector D1 but the setup with two orthogonal polarizers placed in the paths of the MZI (see Fig. 3) does not show interference. For example, one student said "not as many photons will go through. 25% [of the photons will display interference] because only half of the photons going through BS2 will make it through" but he had difficulty with the fact that the quantum eraser setup would show interference and that the setup with two orthogonal polarizers would not. Some students stated that none of the photons would display interference, e.g., "0% [of photons display interference], they are all independent photons." Student responses indicate that some students who may understand the case with two orthogonal polarizers in U and L paths have difficulty with the role of the third polarizer in Fig. 4.

## IV. QuILT DEVELOPMENT

The difficulties discussed above indicate that even after traditional instruction in relevant concepts, upper-level undergraduate and Ph.D. students could benefit from a tutorial-based approach to better learn the concepts involving a single photon propagating through a MZI. Therefore, we developed a QuILT on a MZI with single photons. The QuILT includes a warm-up and a tutorial which strives to help students learn these concepts. It makes use of a computer simulation in which students can manipulate the MZI setup to predict and observe what happens at the detectors for different setups. The QuILT can be used in class to give students an opportunity to work together and check their answers with a partner.

The MZI with single photons QuILT builds on students' prior knowledge and was developed by taking into account the difficulties discussed above. The development of the QuILT was a cyclical, iterative process which included the following stages: 1) development of a preliminary version of the QuILT based on the research on student difficulties; 2) implementation and evaluation of the QuILT by administering it to individual students and measuring its effectiveness via pre-/post-tests; and 3) refinement and modifications based upon the feedback from the implementation and evaluation. The QuILT was also iterated with four faculty members and two Ph.D. students to ensure that the content and wording of the questions are appropriate. We administered the QuILT to several Ph.D. students and upper-level undergraduate students to ensure that the guided approach is effective and the questions were unambiguously interpreted. Modifications were made based upon the feedback.

The first part of the QuILT helps students reason about how a single photon exhibits both the properties of a wave and a particle in different parts of the same experiment, has a non-zero probability of being found in two locations simultaneously (i.e., the photon state is a superposition of path states), and interferes with itself due to the two possible paths through the MZI. Students also are guided to think about how adding or removing optical elements such as beam-splitter 2 or detectors can give "which-path" information about the photon arriving at the detectors D1 or D2 and affect whether interference is observed at the detectors. The second part of the QuILT helps students reason about how polarizers can remove or reinstate interference at the detectors. Checkpoints are also included to help students check their predictions in different situations and reconcile the differences between their predictions and what actually happens. These checkpoints help students understand whether they are reasoning correctly up to a certain point.

The following question is designed to help students reason about the role of beam-splitter 1:
*Consider the following conversation between three students:*
- *Student A: BS1 divides the photon state into two halves. That means that a photon has been divided into two photons with the energy of each photon in the two paths being half of the energy of the photon that entered BS1. If the path difference in the U and L paths of the MZI were set up such that there was an intermediate interference at each detector D1 and D2 (neither fully constructive nor fully destructive), there would be a possibility of both detectors registering a photon at the same time with half the energy of the incoming photon.*
- *Student B: I disagree. Beam-splitter 1 causes the incoming photon state to become a superposition of the two path states U and L, but neither the photon nor its energy is split in half. If the energy was split in half, this would mean that the wavelength of the photon was doubled, which is not the case. Beam-splitter 1 simply makes the single photon state delocalized.*
- *Student C: I agree with Student B's statement. For a single photon, if the MZI was set up such that there was intermediate interference at detectors D1 and D2, only one detector will register a photon, not both. Registering a photon corresponds to a measurement which collapses the state of the photon at the point of detection and localizes it (the photon gets absorbed). We observe interference at the detectors because a single photon interferes with itself.*

*With whom do you agree? You can agree with more than one student. Discuss your preceding answer with a partner and explain your reasoning.*

The following question builds on the preceding question and helps students reason about how a photon can be localized or delocalized depending on the situation:
*Consider the following conversation between three students:*
- *Student A: How can a single photon be in both the U and L paths of the MZI simultaneously if only one detector D1 or D2 clicks and registers a photon? It must go through only one path if only one detector clicks.*
- *Student B: Registering of a photon at the detector corresponds to a measurement of the photon's position via its interaction with the atoms in the detector. The photon is absorbed by the detector during the detection process.*
- *Student C: I agree with Student B's statement. A single photon can be delocalized or localized depending on the situation. For example, the single photon state is delocalized while going through the U and L paths but becomes localized upon detection because measurement collapses the state. Then, the photon gets absorbed by the material in the detector.*

  *With whom do you agree?*
*Discuss your preceding answer with a partner and explain your reasoning.*

In the QuILT, the delayed choice experiment is discussed to help students reason that causality is not violated in this type of an experiment. The delayed choice experiment involves inserting or removing beam-splitter 2 after the photon has already passed through beam-splitter 1. Because the photon is in a superposition state after passing through beam-splitter 1, adding or removing beam-splitter 2 after the photon has already passed through beam-splitter 1 does not cause the photon to "choose" one path or the other; rather, beam-splitter 2 allows both components of the photon path state to be projected in both detectors so that interference is displayed at the detectors. If beam-splitter 2 is removed, the photon is still in a superposition of the path states, but only one component of the photon path state can be projected in each detector and interference is not observed.

For example, the following question and the additional help that follows guides students to reason about how the interference is affected when beam-splitter 2 is removed or inserted after the photon propagates through beam-splitter 1.

*Choose all of the following statements that are true about the case in which the second beam-splitter BS2 is removed:*
  *(I) The point detectors D1 and D2 can only project the superposition state of the photon along the U path state or L path state, respectively.*
  *(II) No interference is observed at either detector and each detector has a 50% probability of registering a photon, regardless of the phase difference between the U and L paths.*
  *(III) It is useless to calculate the phase difference between the photon state due to the U and L paths for information about interference because we have WPI about each photon that arrives at detectors D1 or D2 (because*

*detector D1 can only project the component along the U path and detector D2 can only project the component along the L path).*

Students are also guided to think about how placing additional detectors in the paths of the MZI can destroy the interference observed at the detectors. If an additional detector is placed in one path of the MZI (see Fig. 5), the photon path state collapses to either one path state or the other. After the photon state collapses to either one path state or the other, there is no possibility for interference to occur at the detectors at the end after BS2 because interference is only observed when both path states of the photon can be projected in a detector after beam-splitter 2. The following question helps students reason about the role of an additional detector placed in one of the paths of the MZI:

*Now we will explore how inserting additional photo-detectors in the U and/or L paths can yield information about which path the single photon went through (WPI) and destroy the interference at the detectors placed after BS2. A photo-detector absorbs the photons that it detects.*

*Choose all of the following statements that are correct if you insert an additional detector into the lower path (see Figure 5) and the source emits a large number (N) of single photons.*

*(I) The interference is unchanged (without the phase shifter, N photons reach D1 and no photons reach D2).*
*(II) The interference vanishes.*
*(III) Changing the thickness of the phase shifter will not affect the number of photons reaching detectors D1 and D2.*

*Explain your reasoning for the preceding question.*
*In Figure 5, why will changing the thickness of the phase shifter not affect the number of photons arriving at the detectors? Explain your reasoning below.*

The QuILT also guides students to reason conceptually about how adding polarizers into one or more paths of the MZI can affect the interference observed at the detectors. Since students had difficulty reasoning about how two orthogonal polarizers would eliminate the interference at the detectors, the QuILT strives to help students with this concept. For example, the following question helps students think about the number of photons arriving at the detectors and whether they display interference for the MZI setup with two orthogonal polarizers placed in the two paths of the MZI when the source emits +45° single photons:

*If you place two polarizers with orthogonal polarization axes in the MZI setup (see figure 3) and turn on the +45° polarized single photon source, choose all of the following statements that are correct about what you expect to observe at the detectors after a very large number of photons (N) have been emitted from the source.*
*(I) Interference is displayed and it is identical to that observed with only one of the polarizers present.*
*(II) No interference is displayed.*
*(III) N/4 photons reach detector D1.*
*(IV) Placing a phase shifter in one of the paths and changing its thickness gradually WILL NOT change how many photons arrive at the detectors.*
*Explain your reasoning for the preceding question.*

Many students had difficulty reasoning about the difference between the case in which two orthogonal polarizers are placed in the two paths of the MZI (see Fig. 3) vs. the quantum eraser setup (see Fig. 4). The following series of questions was included in the QuILT to help students reason about the number of photons reaching detectors D1 and D2 and whether interference would be observed with +45° polarized single photons emitted from the source:

*You insert a polarizer with a horizontal polarization axis in the U path and a polarizer with a vertical polarization axis in the L path. You also insert a third polarizer with a +45° polarization axis before detector D1. When you turn on the +45° polarized single photon source, what do you expect to observe at the detectors?*
*(I) No interference is observed at the detectors.*
*(II) Interference is displayed at detector D1.*
*(III) N/4 photons reach detector D1.*
*(IV) Placing a phase shifter in one of the paths and changing its thickness gradually WILL NOT change the number of photons reaching detector D1.*
*Explain your reasoning for the preceding question.*
*Consider the following conversation between three students:*
*Student A: How can you tell that approximately 1/4$^{th}$ of the +45° polarized photons emitted from the source arrive at detector D1 shown in the figure 6 and show interference?*
*Student B: Let me show you a qualitative description of the approximate numbers in a diagram (see Figure. 6).*

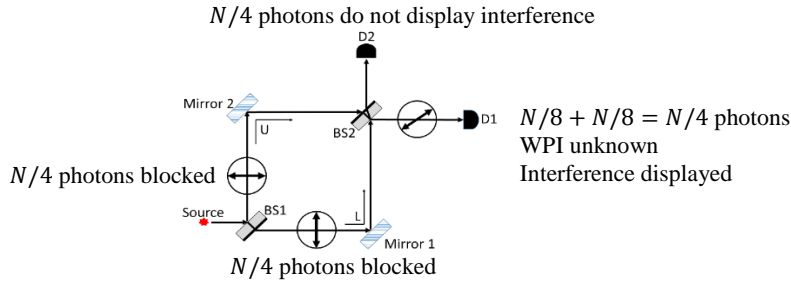

*Figure 6. Qualitative description of the number of photons reaching the detectors in a quantum eraser setup.*

There is interference displayed at detector D1 because the +45° polarizer "erases" the WPI for the photons passing through it. Placing a phase shifter in one of the paths and changing its thickness gradually will change the number of photons reaching detector D1. However, in the setup shown above with no phase shifter, the phase difference between the two paths in the setup is such that the vertically and horizontally polarized components arrive in phase between BS2 and detector D1, resulting in a photon with a +45° polarization component. Thus, no photon is absorbed by the +45° polarizer.

*Student A: Why don't the photons arriving at detector D2 display interference?*
*Student B: Orthogonally polarized beams of light do not interfere, regardless of the phase difference between them.*
*Student C: Yes. The two orthogonally polarized components of the photon state arriving from the two paths between BS2 and D2 cannot interfere. We have WPI for those photons arriving at detector D2 since there is no +45° polarizer between BS2 and detector D2 and thus there will be no interference displayed in the given case when we have two orthogonal polarizers in the U and L paths.*
*Do you agree with Student B and Student C? Explain.*

Students are also given the opportunity to use a computer simulation to check their answers to questions about whether placing additional detectors or polarizers into one or more of the paths of the MZI will affect the interference at detectors D1 and D2. Students can reconcile the differences between their prediction and what they observe using the computer simulation. In the computer simulation, a screen is used in place of point detector D1 and the photon has a transverse Gaussian width as opposed to being a collimated beam having an infinitesimally small transverse width. Students are guided to think about how the transverse Gaussian profile of the photon may yield constructive or destructive interference at different points on the screen, creating an interference pattern on the screen (in situations in which interference should be observed). Students are told that the advantage of the screen (as opposed to point detectors D1 and D2) is that an interference pattern is observed without placing a phase shifter in one of the paths and changing the path length difference between the two paths. For the case with point detectors D1 and D2, the thickness of the phase shifter must be changed in order to observe interference (if interference is displayed in a particular case).

Figure 7 shows a screen shot of the simulation in which an additional detector was placed in one of the paths of the MZI. Students can use the computer simulation to verify that there are no interference fringes on the screen when an additional detector is placed in one of the paths of the MZI (see Fig. 7).

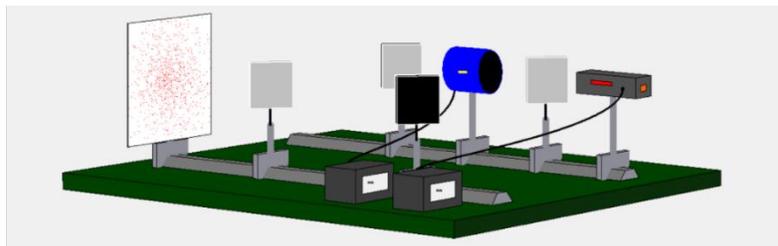

*Figure 7. Screen shot of the computer simulation in which an additional detector (blue device) is placed in one of the paths of the MZI. Simulation developed by Albert Huber.*

In addition, students can reconcile the differences between their predictions and what actually happens when polarizers are inserted into the two paths of the MZI. Students can use the computer simulation to verify that when two orthogonal polarizers are placed in the two paths of the MZI, there is no interference displayed (see Fig. 8).

Students can also use the computer simulation to verify that the quantum eraser setup gives rise to an interference pattern (see Fig. 9).

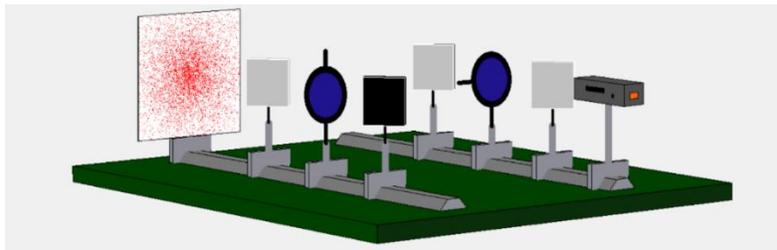

*Figure 8. Computer simulation with polarizers (blue objects) with orthogonal polarization axes placed in the two paths of the MZI. The handle on the polarizer indicates the polarization axis. No interference pattern is observed at the screen.*

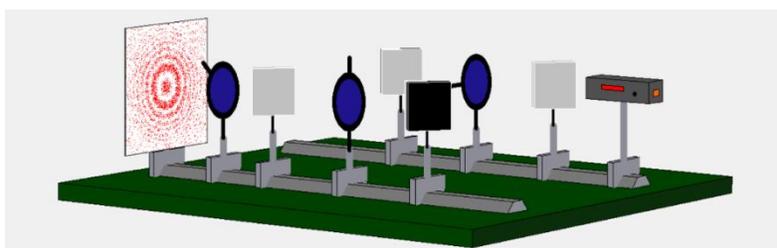

*Figure 9. Computer simulation showing the quantum eraser MZI setup. Interference is observed at the screen.*

After working on the QuILT, students are expected to be able to qualitatively reason about how a single photon can exhibit the properties of both a wave and a particle. They should also be able to describe how a photon can be delocalized or localized depending on the situation and that measurement of a photon's position at the detector collapses the photon path state. Students are also expected to be able to explain the roles of BS1, BS2, and additional detectors placed in the MZI and how these affect the interference at the detectors. Students should also be able to reason about whether a particular MZI setup gives WPI about a single photon and whether inserting a phase shifter will change the number of photons arriving at detectors D1 and D2. In addition, students should be able to reason about how adding polarizers can affect (e.g., eliminate or restore) the interference observed at the detectors. They should also be able to determine whether WPI is known and whether inserting a phase shifter and changing its thickness gradually would affect the number of photons arriving at the detectors in the different cases in which polarizers are inserted into the paths of the MZI.

## V. EVALUATION OF THE QUILT

Once we determined that the QuILT was effective in individual administration, it was administered to students in two upper-level undergraduate quantum mechanics courses ($N = 44$) and Ph.D. students who were simultaneously enrolled in the first semester of a Ph.D. level core quantum mechanics course and a course for training teaching assistants ($N = 45$). First, the students were given a pretest. After the students completed the pretest in class and worked through part of the QuILT in class, they were given one week to work through the rest of the QuILT as homework and were then given a posttest in class. Any students who did not work through the QuILT for any reason were omitted from the posttest data.

The upper-level undergraduate students received full credit for taking the pretest and the tutorial counted as a small portion of their homework grade for the quantum mechanics course. Their posttests were graded for correctness as a quiz for the quantum mechanics course. In addition, the upper-level undergraduates were aware that topics discussed in the tutorial could also appear in future exams since the tutorial was part of the course material for the quantum mechanics course. The Ph.D. students were enrolled in a teaching assistant (TA) training class along with the Ph.D. level core quantum mechanics course. In the TA training class, the Ph.D. students learned about instructional strategies for teaching introductory physics courses. They were asked to work through the QuILT in the TA training class to learn about the effectiveness of the tutorial approach to teaching and learning. They were given credit for completing the pretest, conceptual tutorial, and posttest. However, their actual scores on the posttest did not contribute to the final grade for the TA training class (which was a Pass/Fail course).

Table 1 shows the common difficulties and percentages of students displaying them on the pre/posttest questions and Table 2 displays the average percentage scores on pretest and posttest questions. Questions 1-5 involve concepts related to single photon interference and the role of beam-splitters and additional detectors placed in the path of the MZI. Questions 6-7 involve concepts related to how single photon interference is affected by the addition of polarizers in the paths of the MZI. Part (a) of questions 6 and 7 asks students to compare two different MZI setups with polarizers and describe how they are different, e.g., "You insert a polarizer with a vertical polarization axis in the U path of the MZI and a polarizer with a horizontal polarization axis in the L path of the MZI. Describe what you would observe at D1 and D2 and how this situation will differ from the case in which there are no polarizers present." Part (b) of questions 6 and 7 asks for the percentage of photons that display interference. All questions on the pretest and posttest asked students about single photon interference at point detectors as opposed to detecting screens as shown in the computer simulation.

Average normalized gain [26] is commonly used to determine how much the students learned and takes into account their initial scores on the pretest. It is defined as $\langle g \rangle = (\%\langle S_f \rangle - \%\langle S_i \rangle)/(100 - \%\langle S_i \rangle)$, in which $\langle S_f \rangle$ and $\langle S_i \rangle$ are the final (post) and initial (pre) class averages, respectively [26]. The average normalized gain from pretest to posttest on questions related to difficulties involving interference of light, the wave/particle duality of a single photon, the role of BS2, and the role of additional detectors placed in one of the paths of the MZI, and the role of polarizers placed in the paths of the MZI was 0.78.

*Table 1. Common difficulties and percentages of undergraduate students (UG) and Ph.D. students (G) displaying them on the MZI pretest/posttest questions involving single photons. The number of students who took the pretest does not match the posttest because some students did not finish working through the QuILT and their answers on the posttest were disregarded.*

| Common Difficulty | Pretest UG ($N = 44$) | Pretest G ($N = 45$) | Posttest UG ($N = 38$) | Posttest G ($N = 45$) |
|---|---|---|---|---|
| Q1 Ignoring interference phenomena | 66 | 56 | 21 | 36 |
| Q2 BS1 causes the photon to split into two parts and halves the photon energy | 32 | 24 | 11 | 20 |
| Q2 Photon must take either U or L path | 43 | 36 | 11 | 16 |
| Q3 and Q4 Removing or inserting BS2 does not affect the probability of the detectors D1 and D2 registering photons | 41 | 47 | 16 | 9 |
| Q5 A photo-detector placed in the U or L path may absorb photons but does not affect whether interference is observed if photons arrive at detectors D1 and D2 | 41 | 40 | 0 | 7 |
| Q6 Interference is displayed in an MZI setup with two orthogonal polarizers placed in the two paths (one in each path) | 39 | 47 | 8 | 27 |
| Q7 The quantum eraser setup is not different from placing two orthogonal polarizers in the two paths of the MZI except fewer photons reach the detectors. | 41 | 47 | 5 | 29 |

Question 1 on the pre/posttest assessed student understanding of the classical interference of light in a situation in which a beam of light (instead of single photons) is sent through the MZI. In the first year of administration, 36 students were asked to explain why they agreed or disagreed with the following statement for the basic MZI setup (Fig. 1) without the phase shifter: "If the source produces light with intensity $I$, the intensity of light at each point detector D1 and D2 will be $I/2$ each." In the second year of administration, this question was modified and 53 students were asked to explain why they agreed or disagreed with the following statements for the basic MZI setup (Fig. 1) without a phase shifter: "If the source emits $N$ photons one at a time, the number of photons reaching detectors D1 and D2 will be $N/2$ each." Both statements are incorrect because the MZI setup is such that there is completely constructive interference at D1 and completely destructive interference at D2. Therefore, the light (or single photons) from the U and L paths arrives completely in phase at detector D1 with intensity $I$ ($N$ photons arrive there) and arrives out of phase at D2 and no light (or photon) arrives there. However, Table 1 shows that 66% of the undergraduate students and 56% of the Ph.D. students incorrectly agreed with this statement in the pretest, indicating that they did not take into account the interference phenomenon taking place at the detectors. After working on the QuILT, this difficulty was reduced. Students (undergraduate and Ph.D. students) were given full credit for this question if they stated that they disagreed with the statement and explained that there would be constructive interference at detector D1 and destructive interference at D2.

*Table 2. Average percentage scores on the MZI pretest/posttest for undergraduate students (UG) and Ph.D. students (G). The number of students who took the pretest does not match the posttest because some students did not finish working through the QuILT and their answers on the posttest were disregarded.*

|  |  | Q1 | Q2 | Q3 | Q4 | Q5 | Q6a | Q6b | Q7a | Q7b |
|---|---|---|---|---|---|---|---|---|---|---|
| UG ($N = 44$) | Pretest | 8 | 31 | 18 | 11 | 61 | 17 | 20 | 11 | 14 |
| UG ($N = 38$) | Posttest | 72 | 86 | 87 | 70 | 97 | 85 | 85 | 86 | 85 |
| G ($N = 45$) | Pretest | 21 | 41 | 22 | 13 | 50 | 36 | 40 | 29 | 31 |
| G ($N = 45$) | Posttest | 66 | 76 | 86 | 72 | 87 | 67 | 73 | 67 | 76 |

Question 2 on the pre/posttest assessed students' understanding of the wave nature of a photon. Students were asked to consider the following conversation between two students and explain why they agreed or disagreed with the statements:

*Student 1: "BS1 causes the photon to split in two parts and the energy of the incoming photon is also split in half. Each photon with half the energy travels along the U and L paths of the MZI and produces interference at the detectors."*

*Student 2: "If we send one photon at a time through the MZI, there is no way to observe interference at the detectors. Interference is due to the superposition of waves from the U and L paths. A single photon must choose either the U or L path."*

Neither student is correct because a photon does not split into two parts with half the energy of the incoming photon but a single photon can be in a superposition of the U and L path states. 32% of the undergraduate students and 24% of the Ph.D. students incorrectly agreed with Student 1 in the pretest. After working on the QuILT, Table 1 shows that this difficulty involving the splitting of photons was reduced. Furthermore, 43% of the undergraduate students and 36% of the Ph.D. students incorrectly agreed with Student 2 in Question 2 on the pretest claiming that a photon must take either the U or L path. In the posttest, students performed better. Students who stated that they disagreed with both students and stated correct reasons were given full credit. Some students who agreed with Student 1 (i.e., that the photon is split with half the energy) wrote statements that were partially correct, e.g., "I agree with student 1 because the photon goes into a superposition state and interferes with itself." Students who wrote these types of statements received half credit since the statement that the photon goes into a superposition of path states after BS1 is correct. Students who agreed with Student 2 (i.e., that the photon must choose either the U or L path) were given a score of zero.

Questions 3 and 4 on the pre/posttests evaluated student understanding of the role of BS2. If BS2 is present, it evolves the state of the photon such that both the U and L path components of the photon state can be projected into each detector and the photon interferes with itself at the detectors D1 and D2. In the setup students were given, without the phase shifter in Fig. 1 (when BS2 is present), constructive interference occurs at D1 (the single photons always arrive at D1) and destructive interference occurs at D2 (no photon reaches D2). If BS2 is not present, the photon is still in a superposition of U and L path states after BS1 but only the U path component can be projected in detector D1 and only the L path component can be projected in detector D2. Thus, the photons do not display interference and each detector registers the photons with 50% probability. In the pretest, 41% of the undergraduate students and 47% of the Ph.D. students incorrectly claimed that removing or inserting BS2 will not change the probabilities of the photon arriving at D1 and D2. This high percentage is consistent with the fact that these students did not acknowledge the wave nature and interference effects of single photons in response to other questions as well. Students often explicitly claimed that the photon behaves as a point particle and it would not matter whether BS2 was present or not—each detector would register the photon with 50% probability. Table 1 shows that in the posttest, students performed better. Students were given full credit on these questions if they stated that 1) when BS2 is present, D1 registers all photons and D2 registers zero photons, and 2) when BS2 is removed, D1 registers 50% of the photons and D2 registers 50% of the photons. Students were given half credit if they stated that the probabilities would change depending on whether BS2 was present or missing, but wrote the wrong probabilities. Students were given zero credit if they stated that the probabilities do not change whether BS2 is present or missing.

In Question 5 on the pre/posttests, students were shown a MZI with an additional detector placed in the L path between BS1 and BS2 (see Figure 5). They were then asked to describe how this situation compares to the situation in which no detector is present in the L path as in Figure 1 without the phase shifter. In the situation in which an additional detector is placed in the L path between BS1 and BS2, if the detector does not absorb the photon, the photon path state must collapse to the U path. WPI is known and interference not displayed. Table 1 shows that in the pretest, 41% of the undergraduate students and 40% of the Ph.D. students incorrectly claimed that adding a detector in the L path would not change anything or would cause fewer photons to arrive at detectors D1 and D2 because some photons are absorbed. These students struggled with the fact that the detector in the L path acts as a measurement device and will collapse the photon state of the photons not absorbed by it to the U path state. After working on the QuILT, the difficulty with the effect of an additional detector placed in the L path of the MZI was reduced (see Table 1). Students were given full credit if they stated either that there would be no interference or that there would be $N/4$ photons that reach each of the detectors (as opposed to $N$ photons reaching detector D1 and 0 photons reaching D2) when an additional detector is placed in one of the paths of the MZI.

Question 6 on the pre/posttest evaluated student understanding of the effect of placing two orthogonal polarizers in the two paths of the MZI (see Fig. 3). Students were asked to describe this situation qualitatively and to explain what percentage of photons would display interference. After working through the QuILT, the difficulty with how two orthogonal polarizers affect the interference at the detectors was reduced (see Table 2).

Question 7 on the pre/posttest evaluated student understanding of a quantum eraser (see Fig. 4). The addition of the third polarizer causes both components of the photon path state to be projected in the detector D1, erasing WPI about the photons arriving at D1. As the thickness of the phase shifter is varied, the interference displayed at D1 will change (unlike the setup without polarizer 3). In the pretest, 41% of undergraduate students and 47% of Ph.D. students incorrectly claimed that the quantum eraser setup is not different from the setup with two orthogonal polarizers in the paths of the MZI or that fewer photons would reach detector D1 and/or D2 but otherwise they are the same. After working through the QuILT, this difficulty was reduced (see Table 2).

## VI. SUMMARY

The MZI QuILT focusing on single photon interference uses quantum optics to help students comprehend the wave-particle duality of a single photon, interference of a single photon with itself, how measurement collapses the delocalized superposition state of a single photon, and how polarizers can remove or reinstate interference effects. In fact, many students in the class discussed in the preceding section stated that it was one of their favourite QuILTs. For example one student stated "The [MZI QuILT] was pretty cool because I had no idea what the concept of which path information was before." The evaluations of the QuILT using the pretest and posttest are encouraging.

Since the development of the conceptual QuILT involving a MZI, we have also developed additional QuILTs which strive to help students connect conceptual aspects of the MZI involving single photon inference with mathematical formalism using a two state system when polarizers are not present and a four state system involving photon path states and polarization states [7]. These QuILTs help students connect the qualitative understanding of single photon interference in a MZI with mathematical formalism using a product space for the photon path and polarization states and develop a quantitative understanding of how beam-splitters and polarizers affect interference and measurement outcomes.

## ACKNOWLEDGEMENTS

We thank the U.S. National Science Foundation for awards PHY-0968891 and PHY-1202909. We thank Albert Huber for developing the simulation that we adapted in the QuILT. We also are thankful to various members of the department of physics and astronomy at the University of Pittsburgh (especially R. P. Devaty) and to Andrew Daley and Daniel Oi at the University of Strathclyde, U.K. for helpful conversations and suggestions during the development of the tutorial.